\documentclass{emulateapj}
\usepackage{natbib}

\newcommand\etal{et al.}
\newcommand\ie{i.e.}
\newcommand\eg{e.g.}
\newcommand\msun{{\rm M_\odot}}
\newcommand\kms{\ifmmode{\rm km\ s^{-1}}\else$\rm km\ s^{-1}$\fi}
\def\eps@scaling{1.0}
\newcommand\plotrtwo[2]{{
 \typeout{Plotrtwo included the files #1 #2}
 \centering
 \leavevmode
 \columnwidth=.45\columnwidth
 \includegraphics[angle=270,width={\eps@scaling\columnwidth}]{#1}
 \hfil
 \includegraphics[angle=270,width={\eps@scaling\columnwidth}]{#2}
}}
\newcommand\plotthree[3]{{
 \typeout{Plotthree included the files #1 #2 #3}
 \centering
 \leavevmode
 \columnwidth=.30\columnwidth
 \includegraphics[width={\eps@scaling\columnwidth}]{#1}
 \includegraphics[width={\eps@scaling\columnwidth}]{#2}
 \includegraphics[width={\eps@scaling\columnwidth}]{#3}
}}

\shorttitle{N7626 AND N7619 FROM CHANDRA}
\shortauthors{RANDALL ET AL.}
\slugcomment{Astrophysical Journal, accepted}

\begin{document}

\title{Merging Cold Fronts in the Galaxy Pair NGC~7619 and NGC~7626}

\author{S.\ W.\ Randall\altaffilmark{1}, C.\ Jones\altaffilmark{1},
  R.\ Kraft\altaffilmark{1}, W.\ R.\ Forman\altaffilmark{1}, and E. O'Sullivan\altaffilmark{1}}

\altaffiltext{1}{Harvard-Smithsonian Center for Astrophysics, 60
  Garden St., Cambridge, MA 02138, USA}

\begin{abstract}
We present results from {\it Chandra} observations of the galaxy pair
NGC~7619 and NGC~7626, the two dominant members of the Pegasus group.
The X-ray images show a brightness edge associated with each galaxy,
which we identify as
merger cold fronts.  The edges are sharp, and the axes of symmetry of
the edges are roughly 
anti-parallel, suggesting that these galaxies are
falling towards one another in the plane of the sky.  
The detection of merger cold fronts in each of the two dominant member
galaxies implies a merging subgroup scenario, since the alternative is
that the galaxies are falling into a pre-existing $\sim1$~keV halo
without a dominant galaxy of its own, and such objects are not observed.
We estimate the 3D
velocities from the cold fronts and, using the observed radial
velocities of the galaxies, show that the velocity vectors are indeed
most likely close to the plane of the sky, with a relative velocity
of $\sim1190~\kms$.  The relative velocity is consistent with what is
expected from the infall of two roughly equal mass subgroups whose
total viral mass equals that of the Pegasus group.   
We conclude that the Pegasus cluster is most likely currently forming
from a major
merger of two subgroups, dominated by NGC~7619 and NGC~7626.  
NGC~7626 contains a strong radio source, a core with two symmetric
jets and radio lobes.
Although we find no associated structure in the X-ray surface
brightness map, the temperature map reveals a clump of cool gas just
outside the southern lobe, presumably entrained by the lobe, and
possibly an
extension of cooler gas into the lobe itself.
The jet axis is parallel with the projected direction of motion of NGC~7626
(inferred from the symmetry axis of the merger cold front), and the
southern leading jet is foreshortened as compared to the northern
trailing one, possibly due to the additional ram pressure encountered
by the forward jet.
\end{abstract}


\section{Introduction} \label{sec:intro}

Cluster mergers play a major role in determining the properties of the
intracluster medium (ICM) of galaxy groups and clusters.  They drive,
and aid in the study of, several important physical processes in the
ICM, including shock heating, the generation of relativistic
particles, ram pressure stripping (and hence chemical enrichment of
the ICM), magnetic fields, thermal conduction, and turbulence.  {\it
  Chandra} has revealed the presence of cold fronts associated with
several cluster mergers, where the cold central gas of the merging
cluster forms a contact discontinuity with the ambient ICM.  Cold
fronts have been used extensively to study cluster mergers (for a
review, see Markevitch \& Vikhlinin 2007).  In this paper, we examine
the apparent merging cold fronts in the galaxy pair
NGC~7619/NGC~7626. 

NGC~7619 and NGC~7626 are the two dominant elliptical galaxy members
of the Pegasus galaxy group (U842; Ramella \etal\ 2002).  NGC~7619,
which has a higher stellar velocity dispersion (336~\kms\ vs. 258~\kms\,
Wegner \etal\ 2003), brighter absolute magnitude (−22.94 vs. −22.87;
Lauer \etal\ 2007), and is brighter in X-rays ($4.3 \times
10^{41}$~ergs~s$^{-1}$ vs. $1.2 \times 10^{41}$~ergs~s$^{-1}$;
Fabbiano \etal\ 1992; Burstein \etal\ 1997; O'Sullivan \etal\ 2001),
is presumably the more massive of the two galaxies.  Woods \etal\
(2006) identify NGC~7619 as being paired with NGC~7617, which is a
relatively weak X-ray source.  Both NGC~7619 and NGC~7626 are radio
sources, with NGC~7626 showing symmetric jets/lobes on either side of
a central core (Birkinshaw \& Davies 1985; Hibbard \& Sansom 2003).
NGC~7626 is notable for its unusual optical morphology and stellar
velocity field in the core, which shows kinematically distinct
shell-like regions (Jedrzejewski \& Schechter 1988; Forbes \& Thomson
1992; Balcells \& Carter 1993).  {\it Einstein} observations of this
pair found X-ray emission from a diffuse halo associated with each
galaxy (Fabbiano \etal\ 1992; Burstein \etal\ 1997), with a tail of
asymmetric emission extending SW from NGC~7619.  Subsequent {\it
  ROSAT} PSPC observations confirmed this asymmetry, which has been
interpreted as arising from ram pressure stripping as NGC~7619 falls
into the Pegasus group and interacts with the ICM (Trinchieri \etal\
1997).

More recently, Kim \etal\ (2007; hereafter K07) presented results from
{\it Chandra} and {\it XMM-Newton} observations of NGC~7619.  They
confirmed the presence of the stripped tail, and identified a sharp edge
to the NE in the X-ray surface brightness profile.  Based on this
edge, they found a Mach number of $\sim1$ relative to the ICM.
Abundances and temperatures determined from a spectral analysis of the
tail indicate that this material has indeed been stripped from
NGC~7619, and they interpret these results as directly observed evidence for the
enrichment of the ICM as a result of ram pressure stripping. 

We report here on {\it Chandra} observations of the NGC~7619/NGC~7626 pair.
K07 have previously presented results based on {\it XMM-Newton} and
{\it Chandra} ACIS-S observations only of NGC~7619.  We therefore focus on
{\it Chandra} ACIS-I observations of this pair, and compare our
results for NGC~7619 with those of K07 where appropriate.  We give new
results for NGC~7626, and for its interaction with NGC~7619.  
%In
%particular, we interpret the data as indicating an ongoing major
%merger of two roughly equal mass subgroups, dominated by NGC~7619 and
%NGC~7626, that is responsible for the formation of the Pegasus
%cluster. 
The observations and data reduction techniques are described in
\S~\ref{sec:obs}. The X-ray image is presented in \S~\ref{sec:img},
and results on temperature and abundance structure from spectral
analysis are given in \S~\ref{sec:spec}.
In \S~\ref{sec:discuss}, we give new results for the structure and
velocity of the cold front in NGC~7626, show that NGC~7619 and
NGC~7626 are falling towards one another along an axis that is roughly
in the plane of the sky, and argue that the Pegasus
group is currently forming due to a major merger of two
subgroups. Our results are summarized in \S~\ref{sec:summary}.

We assume a distance to the Pegasus cluster of 53~Mpc, based on results from surface brightness fluctuations for NGC~7619 (Tonry \etal\ 2001), which gives a scale of 0.25 kpc/\arcsec\  for $\Omega_0 = 0.3$,
$\Omega_{\Lambda} = 0.7$, and $H_0 = 70$~\kms~Mpc$^{-1}$.  All
error ranges are 68\% confidence intervals (\ie, 1-$\sigma$), unless otherwise stated.

\section{Observations and Data Reduction} \label{sec:obs}

The NGC~7619/NGC~7626 pair was observed by {\it Chandra} on August 20,
2001, for 27~ksec with the {\it Chandra} CCD Imaging Spectrometer
(ACIS), pointed such that each galaxy was visible on the front-side
illuminated ACIS-I CCD array.  NGC~7619 was again observed by {\it
  Chandra} for 40~ksec on September 24, 2003, this time pointed such
that the galaxy was visible on the back-side illuminated S3 CCD.   
These data were reduced using the method described in Randall \etal\ (2008).
All data were reprocessed from the level 1 events files using the latest
calibration files (as of {\sc CIAO4.0}).  CTI and time-dependent
gain corrections were applied where applicable. {\sc LC\_CLEAN} was
used to remove background
flares\footnote{\url{http://asc.harvard.edu/contrib/maxim/acisbg/}}.
For the earlier ACIS-I observation, time bins that were not
within 3$\sigma$ of the mean were discarded. Unfortunately, the ACIS-S observation was badly contaminated by background flares.  After periods with obvious flares were removed, the 
mean rate was calculated for observed quiescent
time intervals, and {\sc LC\_CLEAN} was re-run, forcing the global mean to
equal the quiescent rate.  The resulting cleaned exposure times were 26.4~ksec and 16.6~ksec for the ACIS-I and ACIS-S observations, respectively.

The emission from NGC~7619, NGC~7626, and the surrounding Pegasus group fills the
image field of view for each observation.  We therefore used the
standard {\sc CALDB\footnote{\url{http://cxc.harvard.edu/caldb/}}}
blank sky background files appropriate for each observation,
normalized to our observations in the 10-12 keV energy band.  To generate exposure maps, we assumed a MEKAL model with $kT = 1$~keV, Galactic absorption, and abundance of 30\%
solar at a redshift $z = 0.0125$, which is consistent with typical results from
detailed spectral fits (see \S~\ref{sec:spec}).  

\section{The X-ray Image} \label{sec:img}

\begin{figure}
\plotone{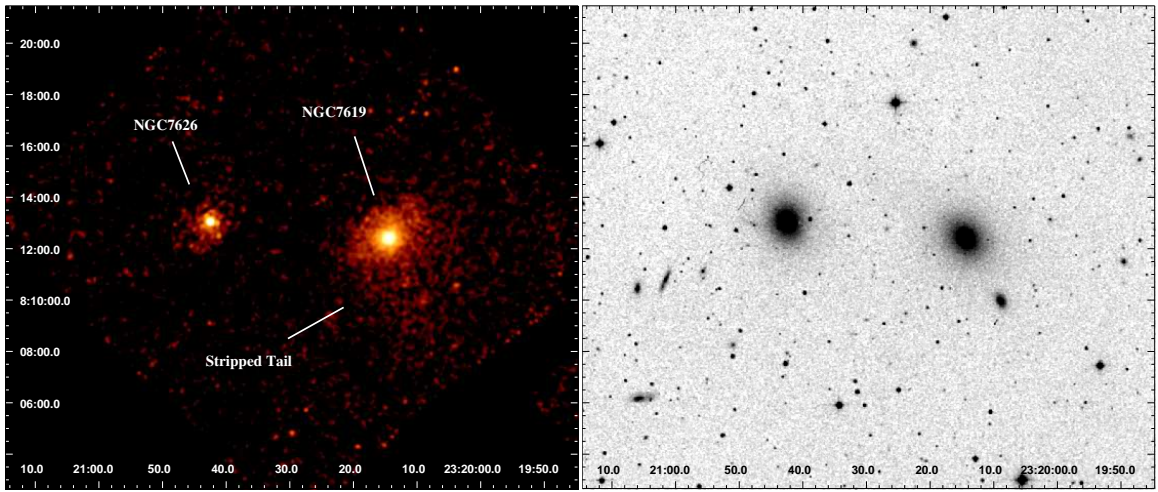}
\caption{
  {\it Left Panel:} Exposure corrected, background subtracted 0.6--5 keV mosaic image of
  the {\it Chandra} ACIS-I and S3 observations of
  NGC~7619 and NGC~7626.  The image has
  been smoothed with an 8\arcsec\ radius gaussian.   For each
  pointing, regions with less than 
  10\% of the total exposure for that observation were omitted.
  Point sources have been removed (see text for details).  NGC~7619
  and NGC~7626 each have a sharp edge in their surface brightness
  profiles, the former to the northeast, the latter to the southwest.
  Additionally, NGC~7619 shows a broad tail of stripped gas to the southwest.
  {\it Right Panel:} {\it DSS} image of the same field.
  \label{fig:fullimg}
}
\end{figure}

\begin{figure}
\plotone{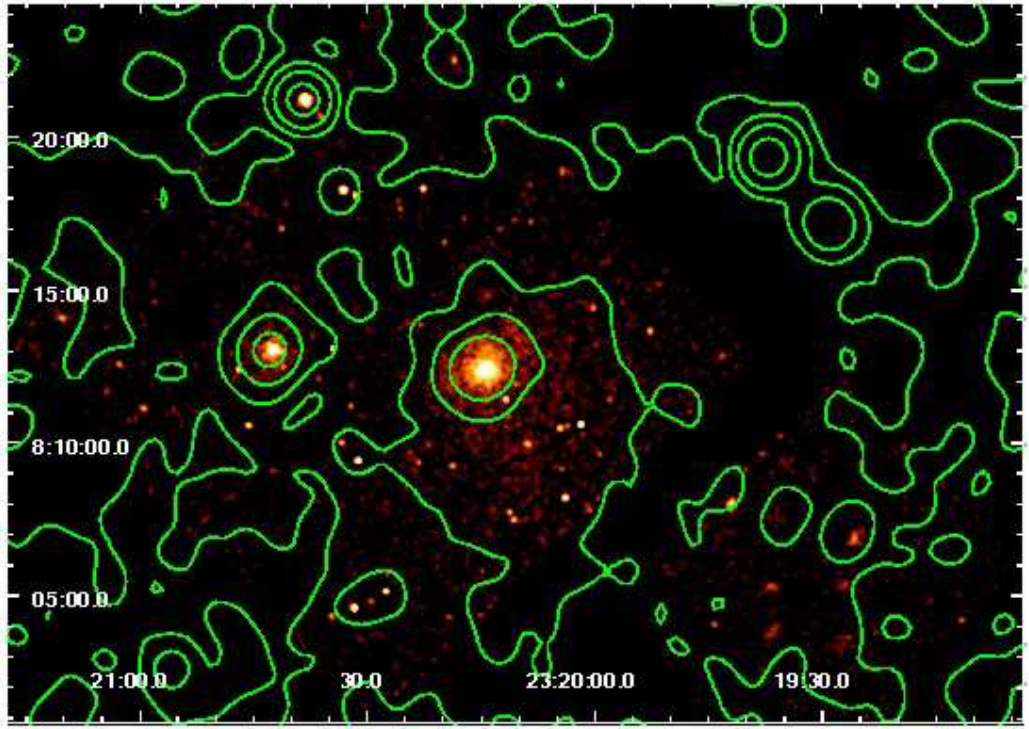}
\caption{
  {\it Chandra} 0.6--5 keV mosaic image, including point sources and smoothed
  with an 8\arcsec\ radius gaussian, with the logarithmic contours
  from a soft band {\it ROSAT} PSPC observation overlain.  The ram
  pressure stripped tail of NGC~7619 is evident, and diffuse emission
  fills the {\it Chandra} field of view.
  \label{fig:rosat}
}
\end{figure}

The exposure corrected, background
subtracted, smoothed mosaic image is shown in
Figure~\ref{fig:fullimg} (the optical {\it DSS} image of the same
field is shown for comparison).  Regions containing point sources were
``filled in'' using a Poisson distribution whose mean was equal to
that of a local annular background region. 
The same image, but with point sources included, is shown in
Figure~\ref{fig:rosat} with {\it ROSAT}
PSPC surface brightness contours overlain to show the extent of the
diffuse emission in the Pegasus group, which fills the {\it Chandra}
field of view.
 In the west, NGC~7619
shows a sharp edge to the northeast, and a diffuse, broad, ram
pressure stripped tail, noted previously (Fabbiano \etal\ 1992;
Trinchieri \etal\ 1997; K07).  In the east, NGC~7626 also shows an
edge in its diffuse emission to the southwest, but in this case there
is no obvious stripped tail.  The symmetry axes of the edges are
roughly anti-parallel, \ie, the cold fronts are directed towards one
another.  The sharpness and relative
orientation of the edges, along with the orientation of NGC~7619's
stripped tail, immediately give the impression of two galaxies falling
towards each other in the plane of the sky.  The edges appear
qualitatively similar to merger cold fronts seen in other systems
(\eg, Markevitch \etal\ 2000; Vikhlinin \etal\ 2001; Machacek \etal\
2005).  We therefore investigate the possibility
that this system shows the early stage of a major merger between two main
subgroups, as opposed to one or more smaller subgroups falling into a
preexisting cluster.

\section{Spectral Analysis} \label{sec:spec}

The X-ray image suggests diffuse emission associated with NGC~7619, NGC~7626, and possibly diffuse emission from the Pegasus group filling the field of view.  We generate a
temperature map as a guide for detailed spectral fitting to
disentangle the various components.  We assume a galactic absorption of
$N_H = 4.82\times 10^{20}$ cm$^{-2}$ throughout.

\subsection{Temperature Map} \label{sec:tmap}

\begin{figure}
\plotone{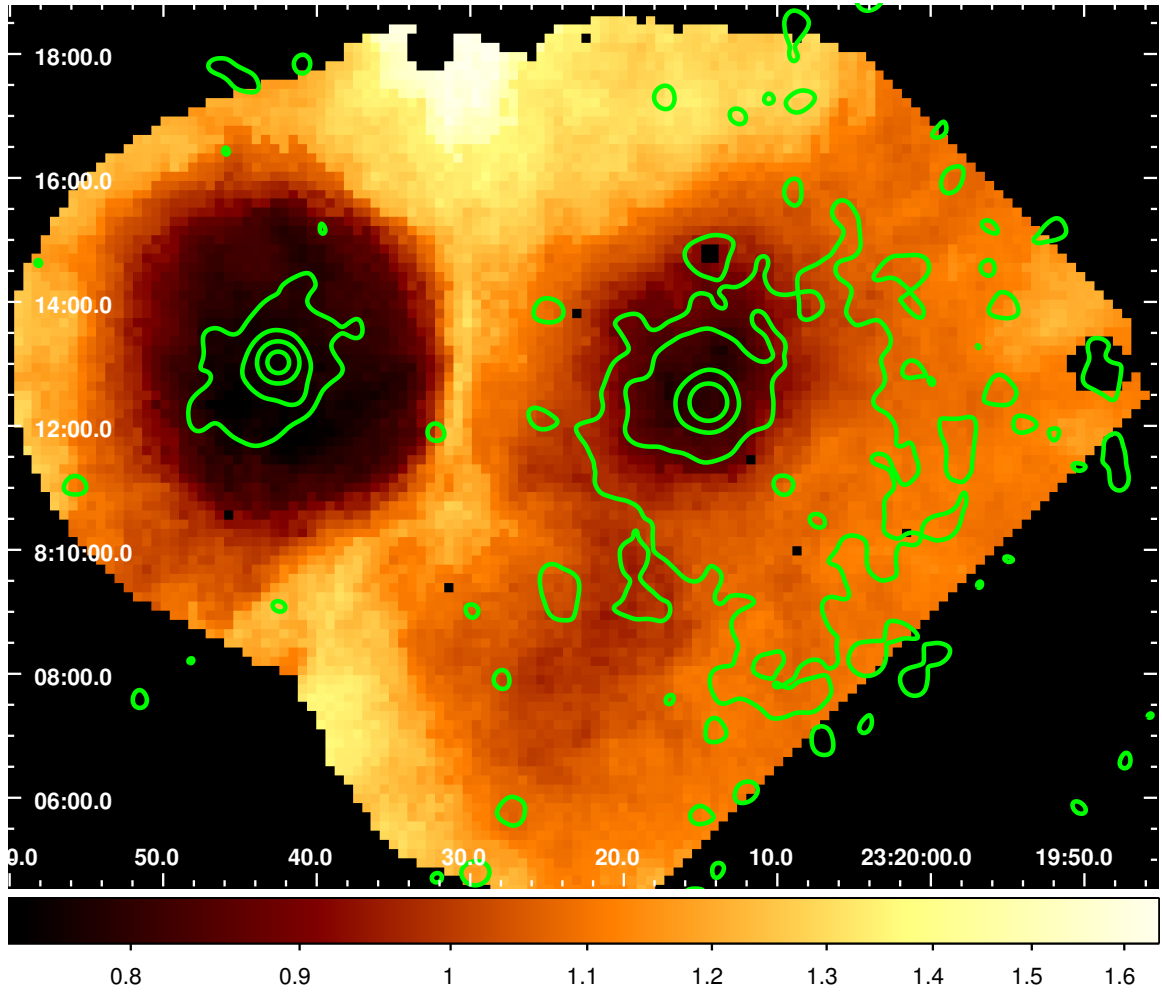}
\caption{Temperature map derived from the ACIS-I observations, with
  {\it Chandra} X-ray logarithmic surface brightness contours overlain
  (with point sources removed). The
  colorbar gives the
temperature in keV.  The holes in the temperature map indicate pixels
that were completely contained within an excluded source region.
\label{fig:tmap}
}
\end{figure}

The temperature map was derived using the same method as developed in
Randall \etal\ (2008), similar to that employed by O'Sullivan et al.\
(2005) and Maughan et al.\ (2006). For each temperature map pixel, we
extracted a spectrum from a circular region containing 1000 net counts
(after subtracting the blank sky background).  The resulting spectrum
was fit in the 0.6 -- 5.0~keV range with an absorbed APEC model using
{\sc XSPEC}, with the abundance allowed to vary.  Maps were generated
using the ACIS-I data only. The 
resulting temperature map is shown in Figure~\ref{fig:tmap}.
Unfortunately, due to the small number of net counts, the extraction
regions for the temperature map pixels were relatively large.  Faint
regions had extraction radii 
on the order of 4.1\arcmin\ (61.5~kpc), while the brightest regions,
near the galaxy cores, had radii of 14\arcsec\ (3.5~kpc).  
%The ACIS-S
%maps had a range in radii of 2.3\arcmin\ (34.5~kpc) to 12\arcsec\
%(3~kpc).  
As a result, each pixel in the temperature maps is highly
correlated with nearby pixels, and the temperature map is effectively
smoothed on large scales. 

Nevertheless, clear structure is visible in the map.  NGC~7619 and
NGC~7626 are each associated with a cloud of cool gas, with central
temperatures of $kT \approx 0.75$~keV.  A cool region extends
southwest from NGC~7619, in the direction of the stripped tail, and
rises to a temperature of about $kT \approx 1$~keV, which is
consistent with findings of K07 using {\it XMM-Newton} data.  Higher
temperature regions, presumably from diffuse group emission and with
$kT \approx 1.2 - 1.3 $~keV, can be seen between the galaxies and
surrounding NGC~7626.  Although the observed surface brightness edges
roughly correspond to regions of transition from cooler to hotter gas,
the temperature maps are too smoothed to match these features in
detail.  An inspection of the error maps suggests that the hot spot
($kT \approx 1.6$~keV) to the north is statistically significant.
While there are bright point sources in this region, most notably two
unidentified sources that were detected as a single blended source in
{\it ROSAT} PSPC observations (Trinchieri \etal\ 1997), these sources
have been adequately masked out when generating the temperature map.
If we remake the map with even larger point source masks the hot spot
remains.  We conclude that, although the fits in this area are
statistically good fits, the hot spot may be an artifact of the method
used to generate the temperature map due to the low number of net
counts and resulting large extraction regions in this area.  More
detailed fits are discussed in \S~\ref{sec:dspec}.

\subsection{Detailed Spectra} \label{sec:dspec}

Based on the derived temperature map and the X-ray image, we
defined 10 regions for detailed spectral analysis.  A summary of the
regions and spectral model for each region are given in
Table~\ref{tab:spectra}.  R1 and R2 cover the central projected 12~kpc
of NGC~7619 and NGC~7626, respectively.  R3 and R4 are just
inside and outside of the surface brightness edge to the northeast in
NGC~7619, while R5 and R6 give the same values for the southwestern
edge in NGC~7626.  R7 is a
featureless region to the north, and should include only group
emission.  Finally, R8, R9, and R10 cover the southern radio lobe
(discussed in \S~\ref{sec:lobe}), a region directly east of the lobe,
and another west of the lobe, respectively. 
The regions are shown overlain on the X-ray image in Figure~\ref{fig:regs}.
The spectrum from each region was initially fit with an
absorbed APEC model over the 0.6-5.0 keV range with the abundance
allowed to vary.
The fits to R1 and R2 showed an excess at hard energies, therefore a power-law component was
included in the fits to these regions to account for the contribution
from unresolved X-ray binaries.  The best-fit photon indicies are are
consistent with $\Gamma \approx 1.5$, as expected for a population of
X-ray binaries (Sarazin \etal\ 2003).  We found that including a
second thermal component, in lieu of the power-law, gave an equally
good fit to the data, though with a second temperature that was higher
than would be expected from the diffuse group emission at
$\sim1.2$~keV.  The results from the multiple thermal component fit
for the core of NGC~7619 are given in Table~\ref{tab:spectra} for
comparison.  We note that this hard energy excess is only detected in
the galaxy cores and its precise nature is not relevant to our main results.

Table~\ref{tab:spectra} shows a clear increase in
temperature across each brightness edge, $0.83^{+0.02}_{-0.01}$ to
$1.05^{+0.09}_{-0.05}$for NGC~7619, and $0.94^{+0.06}_{-0.08}$ to
$1.50^{+0.48}_{-0.27}$ for NGC~7626, with the cooler gas in the
brighter, interior regions.  This clearly establishes these edges as
cold fronts.  The core of NGC~7619 is slightly hotter than that of
NGC~7626, which is consistent with NGC~7619 being the more massive of
the two galaxies.  Using the fits to the detailed spectra, we can derive the approximate mass in 
cool gas in the core of each galaxy.  Assuming a spherical geometry
with a radius of 12~kpc, we find $M_{gas,{\rm N7619}} \approx 1 \times
10^{10}\, \msun$ and  $M_{gas,{\rm N7626}} \approx 5 \times 10^{9}\, \msun$.

\section{Discussion} \label{sec:discuss}

\subsection{Structure of the Cold Fronts} \label{sec:fronts}

\begin{figure}
\plotone{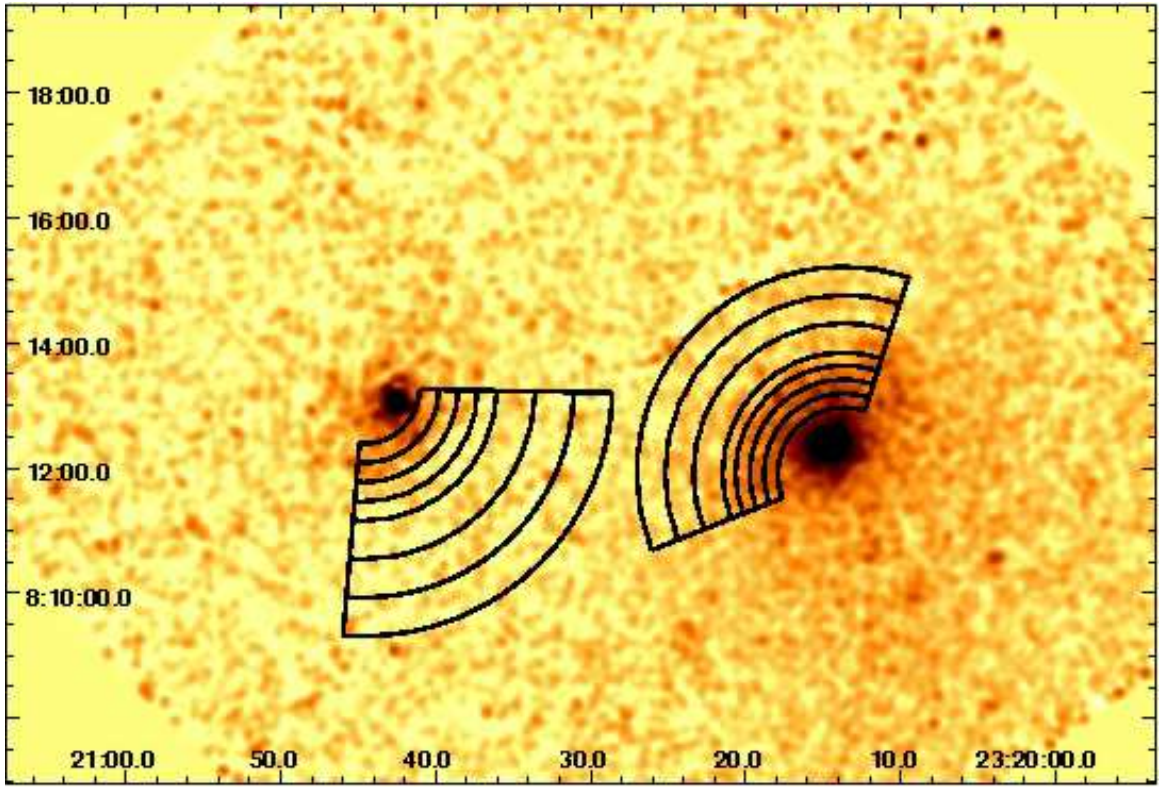}
\caption{
A close up view of the central region of Figure~\ref{fig:fullimg}.
The overlain regions indicate the areas used to generate the projected emission
measure profiles shown in Figure~\ref{fig:profs}.
\label{fig:sect}}
\end{figure}

The {\it Chandra} image (Figure~\ref{fig:fullimg}) shows two sharp
brightness edges, one northeast of NGC~7619, the other southwest of
NGC~7626.  Spectral analysis revealed temperature differences on
either side of the edges, indicating that these edges are cold fronts
(\S~\ref{sec:dspec}, Table~\ref{tab:spectra}).  As such, we expect
there to be a density discontinuity located at each edge.  To measure
the amplitude of these density jumps, we extracted the {\it Chandra}
0.6--5.0 keV surface brightness profile in two sectors, one for
NGC~7619 and one for NGC~7626.  The regions were defined such that the
radii of curvature matched those of the edges.  In neither case did the
position of the center of curvature match that of the optical or X-ray
emission central peak. The extraction regions are shown in
Figure~\ref{fig:sect}.  We use only the ACIS-I data, since the profile
for the edge associated with NGC~7619 has already been accurately
measured and fit by K07 using {\it Chandra} ACIS-S and {\it
  XMM-Newton} data, and we mainly want to cross-check their results
with our own from ACIS-I.  In the case of NGC~7626, only ACIS-I data
are available from {\it Chandra}.

\begin{figure}
\plotrtwo{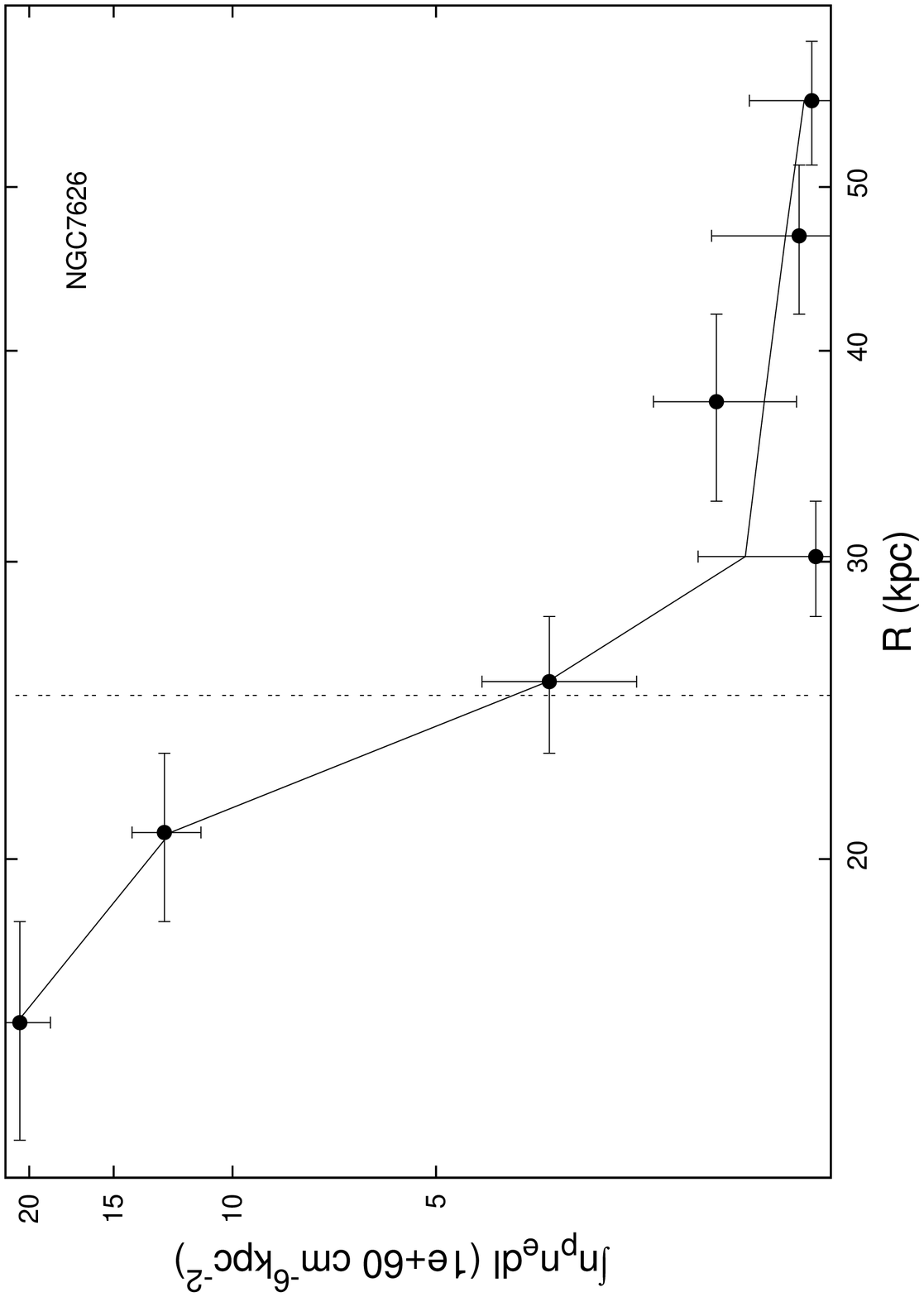}{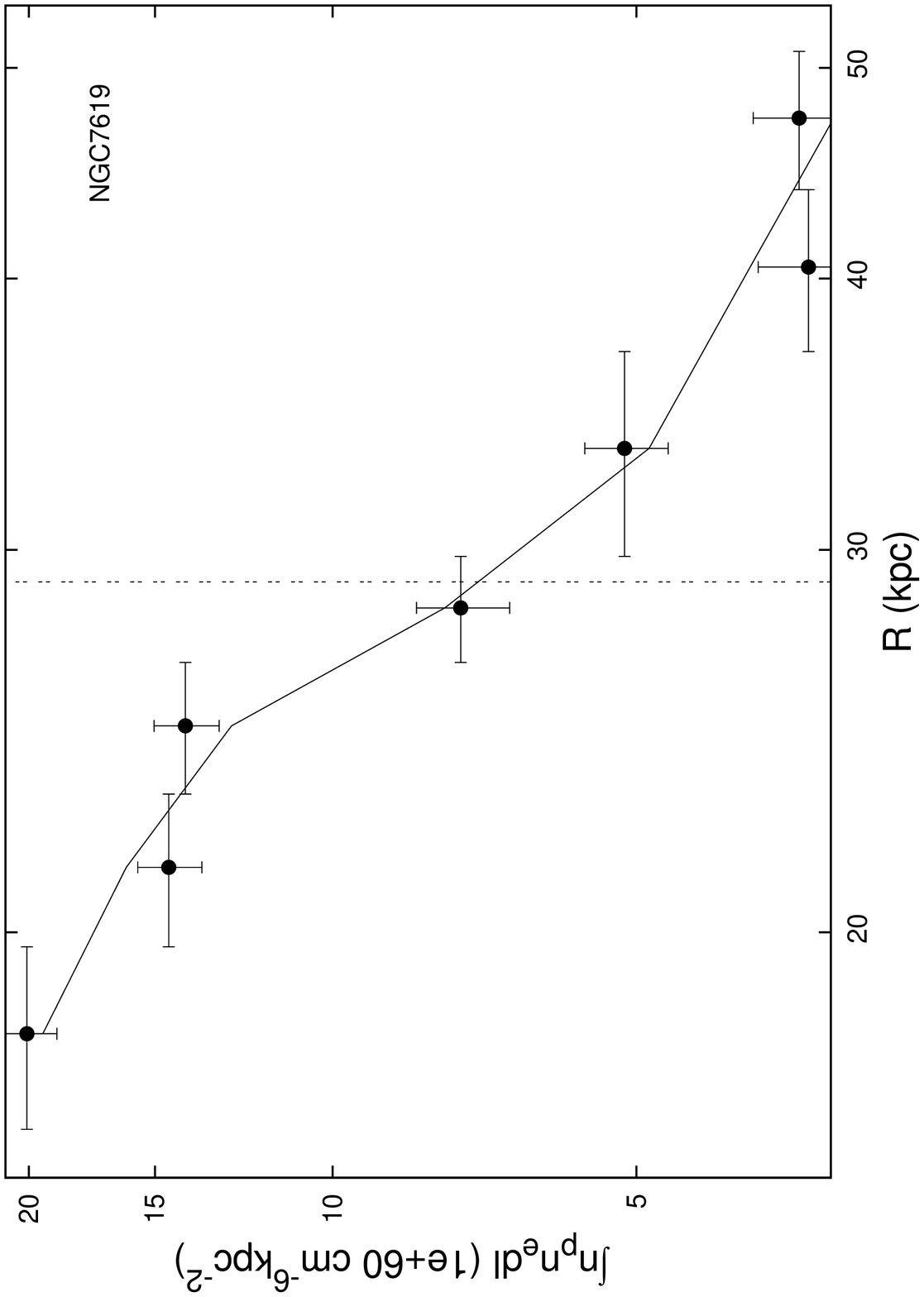}
\caption{
{\it Left Panel:} Integrated emission measure profile across the
surface brightness edge in NGC~7626, extracted from the eastern region
shown in Figure~\ref{fig:sect}.   The x-axis gives
the radius from the apparent center of curvature defined by the
feature.  The best fit two power law density jump model is given by the
solid line. The vertical dashed line shows the best fit location of
the density discontinuity.  {\it Right Panel:} Same for NGC~7619.  Only ACIS-I data
were used in each panel.
\label{fig:profs}}
\end{figure}

The resulting emission measure profiles are shown in
Figure~\ref{fig:profs} (distance is measured from the center of
curvature of the apparent edge).These profiles were fit with a
spherical gas density model consisting of two power laws.  The free
parameters were the normalization, the inner ($\alpha$) and outer
($\beta$) slopes, the position of the density discontinuity ($r_{\rm
  break}$), and the amplitude of the jump ($A$).  The temperature and
abundance for each bin was determined by fitting the projected
emission in coarser bins over the same region, as there were
insufficient counts to perform deprojected
spectral fits.  For the edge in NGC~7619, we find 
$\alpha = -0.45^{+0.23}_{-0.33}$, $\beta = -1.09^{+0.23}_{-0.24}$, 
$r_{\rm break} = 29^{+0.9}_{-1.2}$~kpc, and $A = 2.8^{+1.4}_{-0.2}$.  
%$\alpha = -0.45^{+0.06}_{-0.24}$, $\beta = -1.08^{+0.34}_{-0.38}$, 
%$r_{\rm break} = 29^{+1}_{-2}$~kpc, and $A = 2.7^{+2.1}_{-0.7}$ (90\%
%confidence intervals).  
This is consistent with results from K07, who find a jump in density of
$4.1 \pm 0.6$ (90\% confidence interval).  For the edge in NGC~7626,
we find similar results:  
$\alpha = -0.67^{+0.50}_{-0.21}$,
$r_{\rm break} = 25^{+1.0}_{-1.0}$~kpc, and $A =5.4^{+1.0}_{-0.4}$,
where $\beta$ was fixed at its best-fit value of $\beta = 0.55$.
%$\alpha = -0.80^{+0.09}_{-0.32}$,  $\beta =
%-0.50^{+5.58}_{-0.61}$, $r_{\rm break} = 25^{+2}_{-5}$~kpc, and $A =
%4.6^{+1.3}_{-2.6}$. 
In each case, the two power law model was a 
much better fit to the data than a single power law or beta model.
The best fit models are plotted in Figure~\ref{fig:profs}.

One concern when considering the emission measure profiles is the
contribution from the central galaxy gas, especially in the case of
NGC~7626.  A steep density gradient in the galactic gas will bias the
inner bins high and possibly lead to an artificial increase in the
amplitude of the density jump, particularly if the width of the bins
is large compared to the scale of the system, as is the case here due
to the relatively low photon statistics.  As a test of the
significance of this effect, we fit the emission measure profile of an
identical region to the northwest of NGC~7626, at the same distance
from the galactic center.  In this case, the profile was well described
by a single power-law, with a chi-squared per degree-of-freedom of
$\chi^2_{\nu} = 0.6$ (for comparison, in the southwest a single power-law
model gave $\chi^2_\nu = 4.8$, while the two power-law model gave
$\chi^2_\nu = 0.3$).  The two power-law model did not improve the
chi-squared of the fit, and the amplitude and position of the density
jump could not be well-determined.  This suggests that the central
galactic gas does not strongly bias our measurement of the amplitude
of the jump associated with the edge seen in the southwest.

There are several reasons to identify the edges as
merger cold fronts, as opposed to sloshing fronts as described by
Ascasibar \& Markevitch (2006).  The centers of curvature are not centered on the
optical or X-ray brightness peaks, as should roughly be the case for
sloshing fronts.  The density jumps are large, in the range of a
factor of 2--6,
more characteristic of merger cold fronts than sloshing cold fronts,
which typically have density jumps in the 1--1.5 range (Markevitch \&
Vikhlinin 2007).  The inferred
velocity of the subclumps is mildly supersonic relative to the ambient
ICM (see \S~\ref{sec:geom}).  While this is not impossible for
sloshing fronts, it is more typical of merger fronts.  Although we were
unable to constrain abundance variations across the edges due to
inadequate statistics, K07 found an abundance jump associated with the
edge in NGC~7619 (which has very similar properties to the edge we
observe in NGC~7626), indicating a merger cold front.  Finally,
although the errors are too large to be definitive, we find that the
temperature {\it may} be slightly higher
for the emission just outside the edge in NGC~7626 as compared to a
blank region of sky to the north in the ACIS-I data
(cf.\ R6 and R7 in Table~\ref{tab:spectra}), and to the 
ambient group temperature found by K07 of $kT \approx 1.1$~keV.  This
could indicate a weak shock in this region, though the errors are
large, and if real the difference could be due to normal radial temperature
variations in the ICM (further observations would be required to
detect the presence of a shock front).  We conclude that these
features mostly likely represent merger cold fronts (as opposed to
sloshing fronts), as previously concluded by K07 for NGC~7619.

\subsection{Geometry of the Merger} \label{sec:geom}

Given the apparent sharpness of the cold front edges, and assuming
that these are indeed merger cold fronts, we expect that
the merger axes of NGC~7619 and NGC~7626 are close to the plane of the
sky.  As pointed out by Vikhlinin \etal\ (2001), the 3D velocity of a
merger cold front can be determined by comparing the pressure in the
subclump to the external pressure:

\begin{equation}\label{eqn:pjump1}
\frac{p_0}{p_1} = \left( 1 + \frac{\gamma - 1}{2} M^2_1
\right)^{\frac{\gamma}{\gamma - 1}}, \, \, \, M_1 \le 1,
\end{equation}
\begin{equation}\label{eqn:pjump2}
\frac{p_0}{p_1} = \left( \frac{\gamma + 1}{2}
\right)^{\frac{\gamma+1}{\gamma-1}} M^2_1 \left( \gamma - \frac{\gamma
    - 1}{2 M^2_1} \right)^{\frac{-\gamma}{\gamma - 1}}, \, \, \, M_1 > 1,
\end{equation}
where $p_0$ is the thermal pressure at the stagnation point (which is equal
to the pressure within the subclump), $p_1$ is the thermal pressure in
the external medium, $M_1$ is the Mach number of the subclump relative
to the sound speed in the external medium, and $\gamma = 5/3$ is the
adiabatic index of the gas (Landau \& Lifshitz, 1959).  Using the two
power law density model found in \S~\ref{sec:fronts}, we find a
pressure jump across the cold front in NGC~7626 of
%$p_0/p_1 = 2.9^{+0.98}_{-1.89}$, 
%$p_0/p_1 = 2.9^{+0.6}_{-1.0}$, 
$p_0/p_1 = 3.4^{+0.9}_{-1.2}$, 
which corresponds to a Mach number of 
%$M = 1.3^{+0.2}_{-1.2}$ 
%$M = 1.3^{+0.1}_{-0.4}$ 
$M = 1.4^{+0.2}_{-0.3}$ 
(where the 1-$\sigma$ confidence intervals have been
estimated by separately propagating the upper and lower confidence
intervals on the density jump and temperatures).  
This result
is close to $M \approx 1.2$, found by K07 for the apparently
similar cold front in NGC~7619 (though they point out that if they
take into account the change in abundance across the front they find
$M \approx 0.9$).  For comparison, using the ACIS-I data only (and
accounting for abundance variations), we find 
%$M = 1.0^{+0.08}_{-0.05}$
$M = 1.06^{+0.32}_{-0.10}$
for NGC~7619. 

Using the 3D velocity of the cold fronts in combination with the
observed radial velocity of the galaxies one can estimate the
orientation of the merger axes.  For a sound speed of $c_s = 540$~\kms\
in the ambient gas (at $kT = 1.1$~keV), the 3D velocity of NGC~7626
relative to the ambient ICM is 648~\kms.  The radial velocity is
3405~\kms, -121~\kms\ relative to the mean radial group velocity of 3525~\kms\
(Ramella \etal\ 2002).  Assuming that the radial velocity relative to
the group mean approximates the radial velocity relative to the ICM, we
find that the direction of motion of NGC~7626 is about $10^{\circ}$ from the
plane of the sky.  K07 find that the relative velocity vector of NGC~7619 is
$25-30^{\circ}$ from the plane of the sky, pointing into rather than
out of the plane (its radial velocity of 3820~\kms\ is
larger than the mean group radial velocity).  If we assume that the
projected directions of motion are along the symmetry axes of the cold
fronts, we conclude that NGC~7626 and NGC~7619 are bound to the
Pegasus group (assuming the virial mass of the group to be $8.0 \times
10^{13} \msun$ as given by Ramella \etal\ (2002)) and falling towards
each other with a net relative velocity of approximately -1190~\kms\
(the projected impact parameter is relatively small, $\sim63$~kpc).

The fact that NGC~7626 and NGC~7619, the two dominant members of the
Pegasus group, are falling towards one another with velocities
significantly different (as compared to the group velocity dispersion
of 414~\kms\ given by Ramella \etal\ 2002) from the mean group velocity
suggests that we are seeing a major merger between two similar-sized
groups.  If this is indeed the case, one might expect to see a
double-peaked radial velocity distribution (although the offset of the
peaks may not be large since the merger axis is close to the plane of
the sky).  
Ramella \etal\ (2002) identify 13 members of the Pegasus group.
%The distribution of
%radial velocities for the 13 cluster members identified by Ramella
%\etal\ (2002) is shown Figure~\ref{fig:rdist}.  
Unfortunately, statistical analysis shows that there are an
insufficient number of galaxies to distinguish between a single-peaked
distribution and the bimodal distribution one would expect for two
merging subgroups.

As a further test of this scenario, we estimated the infall velocity
of the merging subgroups.  We assumed that the total virial mass in the
subgroups equals the total virial mass of the Pegasus group, which is
given by Ramella \etal\ (2002) as $8.0 \times 10^{13} \msun$.  The
mass ratio of the subgroups was estimated from the ratio of the square
of the galactic velocity dispersions given by Wegner \etal\ (2003).
This assumes that the group mass scales with the dominant galaxy mass,
which is not true in general, but suffices for our rough estimate here.
This yielded viral masses of $M_{\rm N7619} = 5 \times  10^{13} \msun$
and  $M_{\rm N7626} = 3 \times  10^{13} \msun$.  Each subgroup was
given an NFW potential (Navarro \etal\ 1995) with a concentration
parameter derived from the relations given by Neto \etal\ (2007),
normalized to give the assumed virial mass.  The derived relative
free-fall velocity at the observed separation of 100~kpc was $-960~\kms$,
which is on the order of the $-1190~\kms$ we
derive from cold front measurements.  Relative to the center of mass,
the velocity of 
the smaller subgroup was about 120~\kms\ higher than that of the larger
group (540~\kms\ vs. 420~\kms), similar to what we derive for NGC~7626
and NGC~7619 (648~\kms\ vs. 540~\kms).  While this test is
inconclusive, the fact that the
best-fit point values are in good agreement with our rough estimate
indicates that our model is at least plausible.

\begin{figure}
\plotone{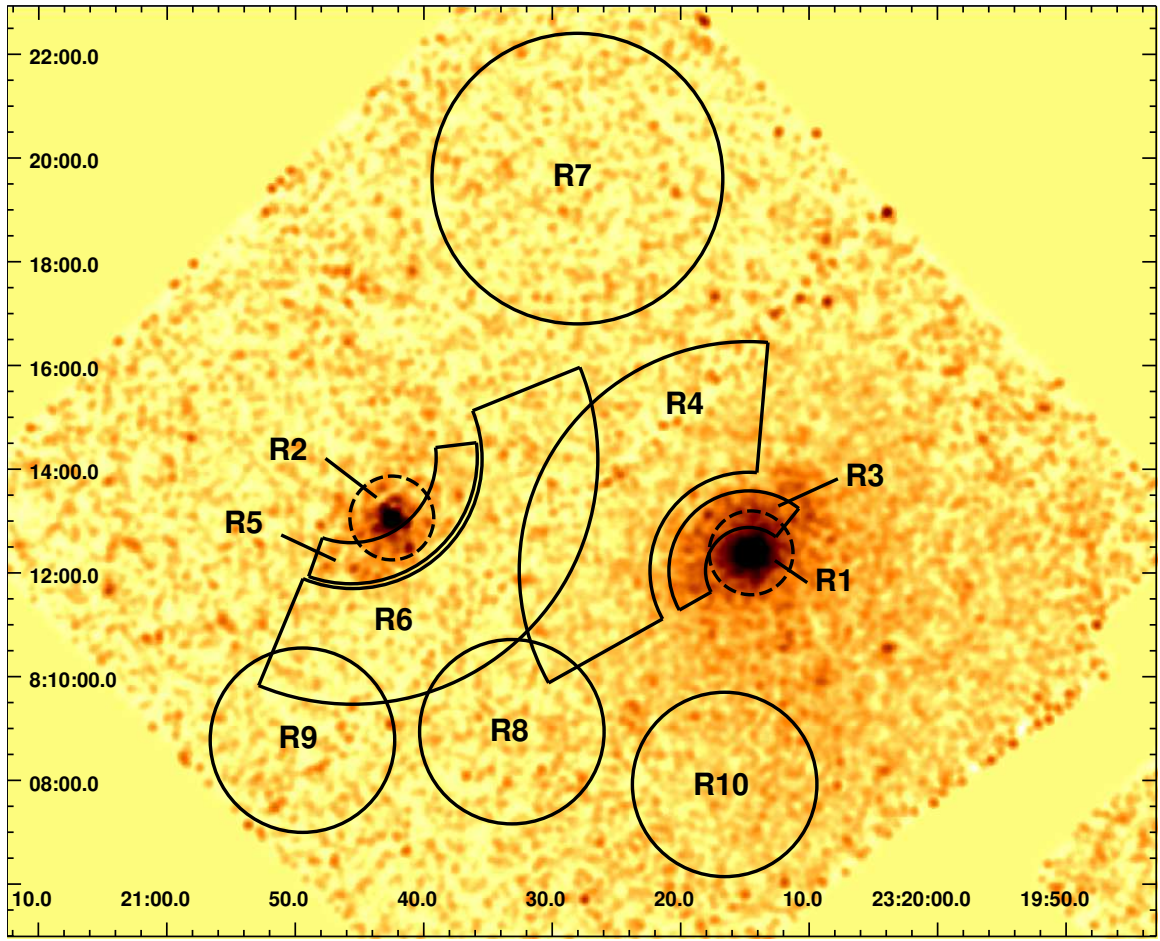}
\caption{
The regions fitted in Table~\ref{tab:spectra} overlain on the 0.6 --
5.0~keV merged ACIS-I and S3 {\it Chandra} image.  The image has been
smoothed with an 8\arcsec\ radius gaussian, and the point sources have
been removed as described in \S~\ref{sec:img} of the text.
\label{fig:regs}
}
\end{figure}

\begin{figure}
\plotone{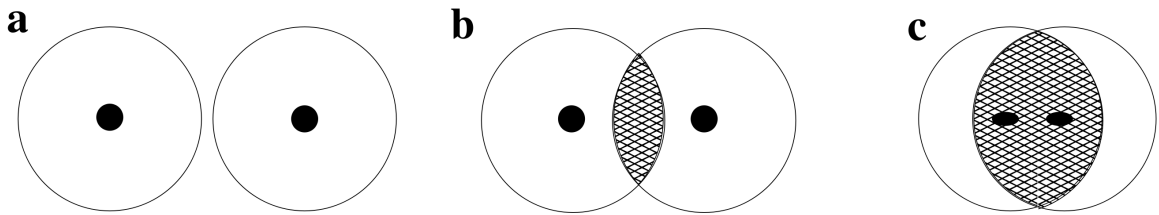}
\caption{
Different stages of a merger between two equal mass subgroups.  The
large circles indicate the virial radii ($\sim0.5$~Mpc) of the
subgroups, while the
filled inner circles indicate the cool gas associated with the central
galaxies.  {\it Panel a:} Initially the groups are falling towards one
another and the virial radii do not overlap (separation is about
1~Mpc). {\it Panel b:} The extended halos of the groups merge
subsonically and form an intermediate region that is at rest with
respect to the center of mass of the system (cross-hatched
region). {\it Panel c:} The relative merger velocity of the subgroups
becomes transonic and the galaxies enter the intermediate region,
become ram pressure stripped, and form merger cold fronts.  The cold
fronts indicate the velocities relative to the merged intermediate ICM.
\label{fig:merge}
}
\end{figure}

This scenario assumes that the gas in the outer regions of the merging
groups has already sufficiently mixed to form an ambient ICM that is
at rest in the center of mass frame, since the cold fronts imply
velocities relative to the ICM that are less than the relative
velocity of the galaxies.  In the case of an isolated
galaxy filled with diffuse 0.75~keV gas that is falling into a group
containing a central massive galaxy, one would expect the velocity
implied by the cold front in the galaxy to equal the relative velocity
of the two galaxies.  In comparison, we find cold front velocities
that are about half the estimated relative velocity of the
galaxies.  A cartoon showing how this can occur for two merging
subgroups is shown in Figure~\ref{fig:merge}.  In panel~{\it a}, the
groups are falling towards one another, but the virial radii have yet
to overlap.  In panel {\it b }, the outer regions of the groups'
extended halos merge subsonically, forming a region in between the
groups that is at rest with respect the the center of mass of the
system and is most likely heated due to adiabatic compression.  As the
groups continue to merge, the relative velocity becomes transonic,
and, as shown in panel~{\it c}, the central galaxies and their
associated cool gas come into contact with the mixed ICM in between
the groups to form ram pressure stripped tails and merger cold fronts.
This scenario predicts a low abundance in the mixed group emission
between NGC~7626 and NGC~7619 (as compared to the abundance in the
ISM and the stripped tail), which is hinted at in our detailed
spectral fits (see Table~\ref{tab:spectra} and Figure~\ref{fig:regs}).

We have suggested that the symmetry axes of the cold fronts indicates
that this is not a head-on merger, with a projected impact parameter
of $\sim63$~kpc (see Figure~\ref{fig:regs}).  In this case, one
expects a curved trajectory for each of the infalling subgroups.
As shown in the left panel of Figure~\ref{fig:fullimg}, and by the {\it
  ROSAT} contours in Figure~\ref{fig:rosat}, the
stripped tail in NGC~7619 is displaced to the east, in the direction
of NGC~7626, relative to the symmetry axis of the cold front,
consistent with our merger scenario.  We do not detect a stripped tail
in NGC~7626, and therefore cannot confirm a curved trajectory.  The
lack of an obvious stripped tail in this system may be due to its
smaller mass and higher velocity relative to ICM (inferred above) as
compared to NGC~7619, resulting in the gas in the extended halo being
stripped away at an earlier stage of the merger. 

Finally, we note that the presence of merger cold fronts in the
two most dominant galaxies in the Pegasus group alone
suggests our merging subgroups scenario.  If we were witnessing, for
instance, NGC~7619 falling into a group dominated by NGC~7626, we
would expect to see the ram pressure stripped tail and merger cold
front in NGC~7619, however we would not expect to find a merger cold
front in NGC~7626 (though we may find a sloshing cold front).  We note
that the existence of the density jump in NGC~7626 is well
determined, with an amplitude of the jump being {\it at least} a
factor of 5 ($1-\sigma$ confidence).  This, along with the temperature
difference across the
edge, which is also well determined, suggests that the brightness edge in
NGC~7626 represents a merger cold front.  The only alternative
scenario consistent with finding two merger cold fronts would be that both
NGC~7626 and NGC~7619 are falling into a
pre-existing cloud of 1~keV gas that had no associated dominant galaxy
of its own.  The authors are unaware of such an object being detected
in the {\it RASS} or other surveys.  We conclude that the data are
fully consistent with the scenario of a major merger between two
subgroups with a small impact parameter.

\subsection{Radio Lobes in NGC~7626} \label{sec:lobe}

\begin{figure}
%\plotthree{radioa.ps}{radiob.ps}{radioc.ps}
\plotone{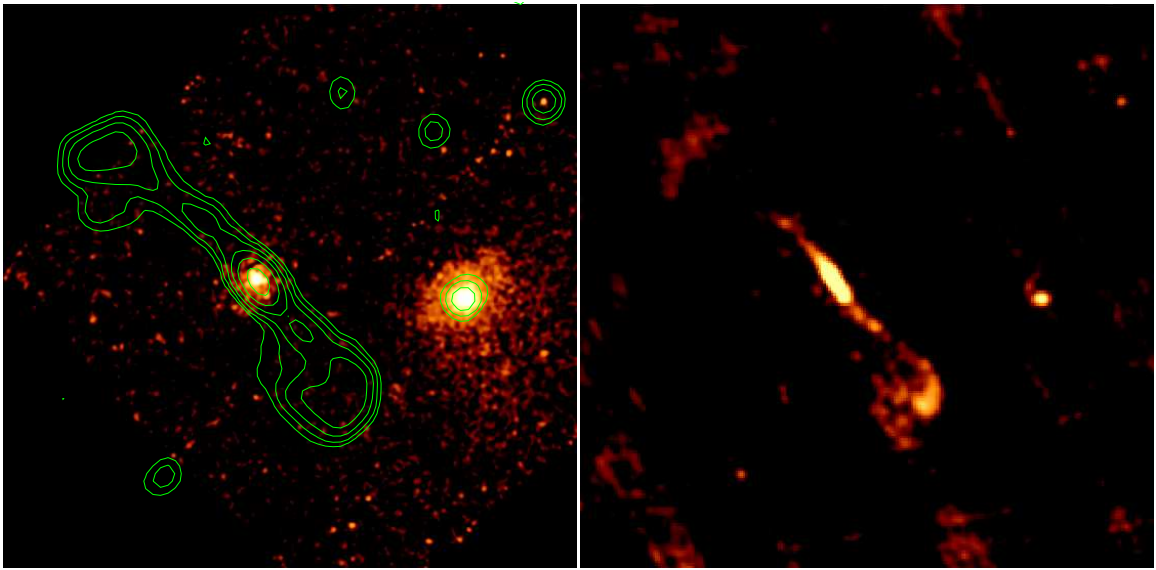}
\caption{
{\it Left Panel:} Close up view of the region near NGC~7626 from
Figure~\ref{fig:fullimg}, with logarithmically spaced NVSS radio
contours overlain. The core 
source, radio jets, and lobes are visible in NGC~7626.    The core source in
NGC~7619 can be seen in the west. {\it Right Panel:} Higher-resolution
20~cm VLA observation of the same region.  The southern lobe clearly
turns back as the jet collides with the ambient ICM.
\label{fig:radio}
}
\end{figure}

\begin{figure}
\plotone{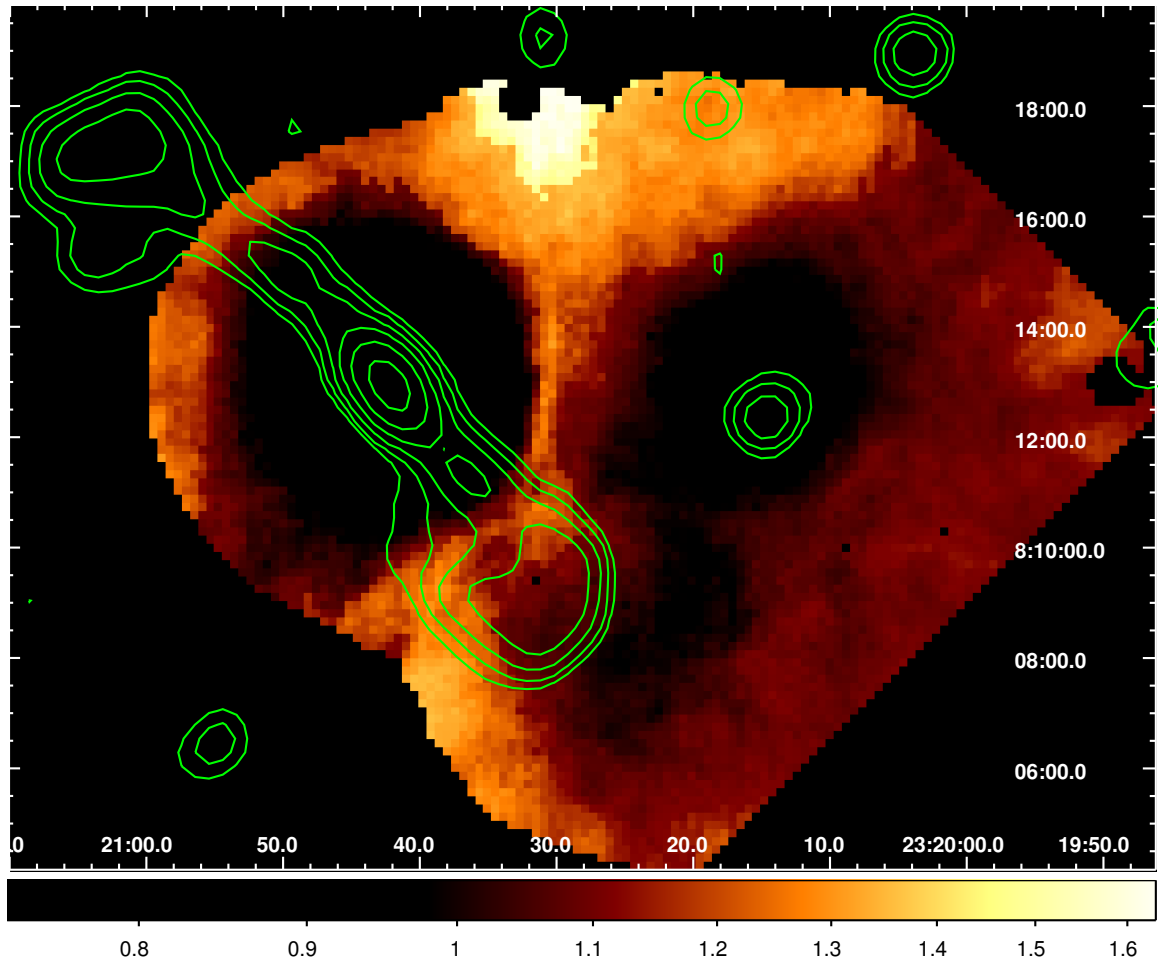}
\caption{
Temperature map with the NVSS radio contours overlain.  The color bar
gives the temperature in keV.  A clump of cool gas can be seen just
outside the southern radio lobe in NGC~7626, the northern edge of which roughly
matches the edge of the lobe.  A tongue of intermediate temperature
($\sim 1.1$~keV) gas extends into the lobe itself.
\label{fig:tmap_rad}
}
\end{figure}

NGC~7626 is a well-known radio source, with symmetric jets on either
side of a central core (Birkinshaw \& Davies 1985; Hibbard \& Sansom
2003).  Figure~\ref{fig:radio} shows VLA 20~cm radio observations of these jets
and their associated lobes along side the {\it Chandra} X-ray
image with the  NRAO VLA Sky Survey (NVSS, Condon \etal\ 1998)
contours overlain.  The compact central radio source in NGC~7619 can
be seen to the west.
There is no
obvious structure in the X-ray surface brightness associated with the southern 
lobe (the northern lobe is outside the field of view).  
Figure~\ref{fig:tmap_rad} shows the temperature map discussed in
\S~\ref{sec:tmap} with the same NVSS radio contours overlain.  A clump
of cool gas can be seen just to the south of the lobe, the northern
boundary of which roughly follows the radio contours.  Furthermore,
there is an extension of cool gas into the core of the lobe,
suggesting the presence of cool gas within the lobe itself.
For a more
detailed comparison, we fit the spectrum of a circular region
encompassing the southern lobe (R8 in Table~\ref{tab:spectra}) and to
two other identical regions just to the east (R9) and west (R10).
R10 contains the cool clump south of the lobe, and some of the ram
pressure stripped tail of NGC~7619.  The
results of these fits are given in Table~\ref{tab:spectra}.
Unfortunately, there are too few counts in these regions to constrain
temperature differences, and we are left with the variations suggested
by the overlapping regions used to generate the temperature map.

As shown in Figure~\ref{fig:radio}, the
jets are roughly aligned with the axis of symmetry of the cold front,
\ie, the presumed projected direction of motion.  As is most clearly
seen from the contours in the left panel of Figure~\ref{fig:radio},
the jets are not completely symmetric, with the
southern jet being about 1.7\arcmin (25~kpc) shorter than the northern jet
(5.78\arcmin\ (87~kpc) vs. 7.46\arcmin\ (112~kpc)).  
Although this does not necessarily imply our merger scenario, we note
that it is at least consistent, in that the foreshortening of the
southern jet may be due to the additional
ram pressure force this jet sees as it pushes into the ICM in the
direction of motion.
If we extrapolate our density model for NGC~7626 given in
\S~\ref{sec:fronts} (and shown in Figure~\ref{fig:profs}) to the tip
of the southern lobe, we find the ram pressure at this point to be
$\sim3 \times 10^{-12}$~dyne~cm$^{-2}$ for an object moving with NGC~7626 at
648~\kms\ (derived in \S~\ref{sec:geom}).  For an outward moving lobe,
the ram pressure on the lobe would be larger due to the additional
velocity of the lobe relative to the ICM.
Alternatively, the jet seems to extend to the edge of the ram pressure stripped
tail of NGC~7619, and may be turned back by the cooler, more dense gas
there.  
The cool clump of the gas south of the lobe could either be material
that has been pushed out of NGC~7626 by the radio lobe, or entrained
gas from the ram pressure stripped tail of NGC~7619.
A deeper X-ray observation of this region (and the region near the
northern lobe) to examine the
interaction of the radio jets and the ambient ICM would be of interest.

\section{Summary} \label{sec:summary}

We interpreted {\it Chandra} observations of the Pegasus group in the
context of a major 
merger of two subgroups, dominated by NGC~7619 and NGC~7626.  Sharp
surface brightness edges in each galaxy are suggestive of merger (as
opposed to sloshing)
cold fronts, due
the the temperature changes across the edges, the high amplitude of
the associated density jumps, the large inferred velocities, and the
offsets between the centers of curvature of the features and the X-ray
brightness peaks.  K07 previously reached a similar conclusion for the
edge in NGC~7619.  The axes of symmetry of the fronts are roughly
anti-parallel, and each velocity vector is found to be close to the
plane of the sky, implying that these systems are infalling towards
one another.  The relative velocity of the galaxies is consistent with
what is expected for two roughly equal mass subgroups in free fall
with a zero impact parameter.  
Although the 3D velocities derived from
the cold fronts are uncertain, the detection of the cold fronts
(\ie, the temperature and density jumps across the edges) is not.
If these are indeed merger cold fronts, the presence of these fronts
alone would suggest a merging subgroup scenario, since the alternative
would be two galaxies falling into a pre-existing 1~keV cloud that did
not contain a dominant galaxy.  Such objects are not observed.
The temperature map shows a clump of cool gas that is apparently
entrained by the southern lobe, and some cooler gas within the lobe
itself.  The cool clump may have been pushed out of NGC~7626 by the
jet, or formed by pile up when the lobe encountered the ram pressure stripped
tail of NGC~7619.
Finally, the southern radio jet in NGC~7626 is foreshortened, which
may be due to 
the additional ram pressure from the merger, or to encountering the
denser material in the 
ram pressure stripped tail of NGC~7619, as it interacts with the
ambient ICM. 

All the data discussed above are fully consistent with our
interpretation of a major merger currently forming the Pegasus
group. Other interpretations may also be consistent with the data,
e.g., NGC~7626 merging into an existing group whose dominant central
galaxy NGC~7619 is undergoing large "sloshing motions" as a result of
the merger. For the reasons detailed above, e.g., the large amplitude
of the density jumps associated with the cold fronts, we argue that
the double merger is more likely. Further observations, that would
provide more accurate spectral data allowing more detailed and precise
hydrodynamic models, may be able to definitively discriminate among the
alternatives.  In addition, since major group mergers may show more
extreme features in the gas relative to cluster mergers (e.g. larger
pressure jumps, larger relative displacements between the central
galaxy and any gas, etc.), more simulations of group mergers would be
useful.

\acknowledgments

The financial support for this
work was partially provided for by the NASA {\it Chandra} grant
GO0-1026X, the Chandra X-ray Center through NAS8-03060, 
and the Smithsonian Institution. 
We thank Pasquale Mazzotta for useful comments.

\clearpage

\clearpage
\begin{deluxetable}{lcccccc}
\tablewidth{7.0truein}
\tablecaption{Spectral Fits \label{tab:spectra}}
\tablehead{
\colhead{Region \#}&
\colhead{CCDs}&
\colhead{$kT$}&
\colhead{Abund.}&
\colhead{$\Gamma$}&
\colhead{$\chi^2$/dof}&
\colhead{Net Cnts.}\\
\colhead{}&
\colhead{}&
\colhead{(keV)}&
\colhead{(solar)}&
\colhead{}&
\colhead{}&
\colhead{}
}
\startdata
R1&ACIS-I/S3&0.778$^{+0.008}_{-0.008}$&0.77$^{+0.90}_{-0.22}$&1.81$^{+0.38}_{-0.52}$&112/90=1.33&4938\\
R1&ACIS-I/S3&0.774$^{+0.008}_{-0.010}$/3.34$^{+2.41}_{-0.88}$&0.76$^{+0.54}_{-0.12}$&&109/83=1.32&4938\\
R2&ACIS-I&0.630$^{+0.031}_{-0.028}$&0.16$^{+3.96}_{-0.07}$&1.09$^{+0.99}_{-1.08}$&15/19=0.79&1282\\
R3&ACIS-I/S3&0.831$^{+0.014}_{-0.015}$&0.32$^{+0.09}_{-0.07}$&&44/36=1.22&1633\\
R4&ACIS-I/S3&1.050$^{+0.092}_{-0.047}$&0.09$^{+0.03}_{-0.02}$&&79/54=1.58&1235\\
R5&ACIS-I&0.938$^{+0.066}_{-0.074}$&0.10$^{+0.05}_{-0.04}$&&4/9=0.42&395\\
R6&ACIS-I&1.500$^{+0.486}_{-0.269}$&0.16$^{+0.52}_{-0.13}$&&16/22=0.73&352\\
R7&ACIS-I&1.263$^{+0.351}_{-0.225}$&0.11$^{+0.16}_{-0.08}$&&9/9=1.00&365\\
R8&ACIS-I&1.358$^{+0.178}_{-0.085}$&0.55$^{+0.75}_{-0.24}$&&15/12=1.24&260\\
R9&ACIS-I&1.116$^{+0.531}_{-0.457}$&0.04$^{+0.24}_{-0.04}$&&6/8=0.75&148\\
R10&ACIS-I&1.068$^{+0.125}_{-0.062}$&0.24$^{+0.22}_{-0.08}$&&3/7=0.45&493\\
\enddata
\end{deluxetable}

% FIGURES

%\begin{figure}[h]
%\begin{center}
%\hbox{
%\includegraphics[angle=270, width=3.5in]{figure6c.ps}
%}
%\caption{Test histogram plot of figure 6a.
%\label{fig:test}}
%\end{center}
%\end{figure}

%\begin{figure}
%\plotone{radvel.ps}
%\caption{
%Radial velocity distribution of the 13 Pegasus group member galaxies
%identified by Ramella \etal\ (2002) relative to the mean group
%velocity.
%\label{fig:rdist}
%}
%\end{figure}

\end{document}